\documentclass[twocolumn,showpacs,preprintnumbers,amsmath,amssymb]{revtex4}

\usepackage{graphicx}% Include figure files
\usepackage{dcolumn}% Align table columns on decimal point
\usepackage{bm}% bold math

\begin{document}

\title{\bf Multi-terminal Spin Transport: Non applicability of linear response and Equilibrium spin currents}
\author{T.~P.~Pareek}
\affiliation{Harish Chandra Research Institute \\ Chhatnag Road, Jhusi,
Allahabad - 211019, India
}%

\begin{abstract}
We present generalized scattering theory for multi-terminal spin transport in systems with broken SU(2) symmetry either due to spin-orbit interaction,magnetic impurities or magnetic leads.
We derive equation for spin current consistent with charge conservation.
It is shown that resulting spin current equations can not be expressed as difference of potential pointing to {\it non applicability of linear response for spin currents} and as a consequence equilibrium spin currents(ESC) in the leads are non zero.
We illustrate the theory by calculating ESC in two terminal normal system
in presence of Rashba spin orbit coupling and show that it leads to spin rectification
consistent with the non linear nature of spin transport.
\end{abstract}

\maketitle
Spin transport has emerged as an important subfield of research in bulk condensed matter system
as well in mesoscopic and nano system\cite{zutic}. In macroscopic systems,
the very definition of spin currents is still debated due to non conservation of spin in presence of SO interaction\cite{rashba} . On the other hand in 
mesoscopic hybrid system 
since current is defined in the leads where SO coupling is absent,therefore, it has been assumed that Landauer-B\"uttiker formula for charge current\cite{buttiker46} (Eq.~(\ref{chcurr}) in this manuscript), which determines current in leads in terms
of applied voltage difference multiplied by total transmission probability, can be straight away generalized for spin currents by replacing total transmission probability with some particular combination of spin resolved transmission probabilities\cite{land-but,jiang,schied}. This simple generalization has been widely used in the literature to study spin dependent phenomena in nanosystems\cite{land-but,jiang,schied}.
The Landauer-B\"uttiker formula in its widely used
form has inbuilt current conservation ,{\it i.e.}, total current is divergence less (div {\bf j}=0)
which follows from basic Maxwell equations of electrodynamics. Physically this implies that total current has neither sources nor sinks this would be true for spin currents as well if
spin is conserved. However, in presence of spin-orbit interaction, magnetic impurities or non-collinear magnetization in leads, spin is no longer a conserved quantity, hence the spin currents can not be divergence less.
Therfore a straight forward generalization of Landauer-B\"uttiker charge current formula to spin currents can not be correct for spin non-conserving systems.

In view of the above discussion in this work we develop a consistent scattering theoretic formulation of coupled spin and charge transport in multiterminal systems with broken SU(2) symmetry in spin space following B\"uttiker's work on charge transport \cite{buttiker46}. The SU(2) symmetry in spin space can be broken due to either SO interaction, magnetic impurities or non-collinear magnetization in leads\cite{tribhu1,pareek}.
Our analysis provides a correct 
generalization of Landauer-B\"uttiker theory for spin transport. In particular we derive a spin currents equation (Eq.~(\ref{spchf}) in this manuscript) consistent with charge current conservation. However, 
the resulting spin current equation can not be cast in terms of spin resolved transmission and reflection probabilities multiplied by voltage difference as is the case for charge current(Eq.~(\ref{chcurr}) in this manuscript)\cite{buttiker46}. Therefore, equilibrium spin currents are generically non zero. {\it Moreover, it implies that linear response theory
with respect to electric field is not applicable to the spin currents (equilibrium as well non-equilibrium).Thus spin currents are intrinsically non-linear in electrical circuits.}
However, this is not surprising since linear response is valid for thermodynamically conjugate variable. In an electrical circuit thermodynamically conjugate variable to electric field is charge current not the spin currents.
We illustrate the theory by calculating
ESC analytically for two-dimensional electron system with
Rashba SO interaction in contact with two unpolarized metallic contacts. Our analytical formula for ESC clearly demonstrates that it is transfer of angular momentum per unit time from SO coupled sample to leads where SO interaction is zero. Therefore,it is truly a transport current in contrast to equilibrium spin currents in macroscopic Rashba medium(see ref.\cite{rashba,silsbee}).

To formulate scattering theory for spin transport we consider a
mesoscopic conductor with broken SU(2) symmetry in spin space connected to a number of ideal magnetic and non-magnetic leads (without SO interaction) which in turn are connected to electron reservoirs.
To include the effect of broken SU(2) symmetry,
it is necessary to write spin scattering state
in each lead along local spin quantization axis.
For magnetic leads local magnetization direction provides a natural spin quantization axis which
we denote in a particular lead $\alpha$ by $\hat{\bm m}_{\alpha}(\vartheta_{\alpha},\varphi_{\alpha})$ where $\vartheta_{\alpha}$ and $\varphi_{\alpha}$
is polar and azimuthal angle respectively.
For nonmagnetic leads 
since there is no preferred spin quantization axis, hence we choose an arbitrary spin
quantization axis $\hat{\bm u}(\theta,\phi)$ which is same for all nonmagnetic leads.
Thus the most general spin scattering state in lead $\alpha$ which can be either magnetic or nonmagnetic is given by,
\begin{eqnarray}
\hat{\Psi}_{\alpha}^{\sigma}(\bm r,t)=\int dE \sum_{n=1}^{N_{\alpha}^{\sigma}(E)}
\frac{\Phi_{\alpha n}(r_{\perp}){\chi_{\alpha}}(\sigma)}{\sqrt{2\pi\hbar v_{\alpha n}^{\sigma}(E)}}  \nonumber \\(a_{\alpha n}^{\sigma}(E)e^{i k_{\alpha n}^{\sigma}(E) x} +b_{\alpha n}^{\sigma}(E) e^{-i k_{\alpha n}^{\sigma}(E)x})
\label{field}
\end{eqnarray}
\noindent where $\Phi_{\alpha n}(r_{\perp})$ is transverse wavefunction of channel $n$ and $\chi_{\alpha}(\sigma)$ is corresponding spin wave function along chosen spin quantization axis, $\hat{\bm u}$ or $\hat{\bm m_{\alpha}}$ such that $\bm S\cdot \hat{\bm u} \chi(\sigma)$=$(\sigma\hbar/2)\chi(\sigma)$  or $\bm S\cdot \hat{\bm m_{\alpha}}\varphi(\sigma)$=$(\sigma\hbar/2)\varphi(\sigma)$ with $\sigma=\sigma$ ( $\sigma$=$\pm 1$,representing local up or down spin components) for nonmagnetic and magnetic leads respectively. Here $\bm S = (\hbar/2)\bm\sigma$ is a vector
of Pauli spin matrices and $N_{\alpha}^{\sigma}$ is number of channels with spin $\sigma$ in lead $\alpha$.
The relation between spin dependent wavevector $k_{\alpha n}^{\sigma}(E)$ and energy $E$ is specified by, $E=\left[\hbar^2 k_{\alpha n \sigma}^2/2m+\varepsilon_{\alpha n}+\sigma\ \Delta_{\alpha}\right]$, where $\varepsilon_{\alpha n}$ is energy due to transverse motion, $\Delta_{\alpha}$ is stoner exchange splitting in the magnetic lead $\alpha$. The stoner exchange splitting is zero for nonmagnetic leads.
The operators $a_{\alpha n}^{\sigma}$ and $b_{\alpha n}^{\sigma}$ are annihilation operator for incoming and outgoing spin channels in lead $\alpha$ and are related via the scattering matrix,
\begin{equation}
b_{\alpha m}^{\sigma}=\sum_{\beta n \sigma^{\prime}}S_{\alpha  m ;\beta n}^{\sigma \sigma^{\prime}}
a_{\beta n}^{\sigma^{\prime}}
\label{scmat}
\end{equation}
The scattering matrix elements $S_{\alpha  m ;\beta n}^{\sigma \sigma^{\prime}}$ provides scattering amplitude between spin channel $n\sigma^{\prime}$ in lead $\beta$ to spin channel $m \sigma$ in lead $\alpha$. These scattering matrix elements will be function of energy E as well
angles, $\vartheta$ and $\varphi$. The angular dependence of scattering matrix elements on polar and azimuthal angle arise due to broken SU(2) symmetry.
Note that for noncollinear magnetization in leads and in absence of SO interaction and 
magnetic impurities, the angular dependence is purely of geometric origin and 
is related to the angular variation of various magnetoresistance phenomena\cite{bauer}.

The current in spin channel $\sigma$ along longitudinal direction $\hat{\bm x}$(through a cross section of lead $\alpha$) and the local
spin quantization axis $\hat{\bm u}$ is defined as,
\begin{equation}
\hat{I}_{\alpha\,\hat{{\bm u}}}^{\hat{\bm x}\sigma}(t)=\frac{\hbar}{2mi}\int\left[\hat{\Psi}^{\dagger \, \sigma}_{\alpha} ({\bm S} \cdot \hat{\bm u})\, \nabla_{x}\hat{\Psi}^{\sigma}_{\alpha} - \nabla_{x}\hat{\Psi}^{\dagger \,\sigma}_{\alpha}\, ({\bm S} \cdot \hat{\bm u}) \hat{\Psi}^{\sigma}_{\alpha} \right] dr_{\perp}.
\label{spcurr1}
\end{equation}
Substituting for $\hat{\Psi}_{\alpha}^{\sigma}$ from Eq.(\ref{field}) into Eq.(\ref{spcurr1}, we get an expression for spin current in terms of creation and annihilation operators. On the resulting expression we perform quantum statistical averages and after a lengthy algebra we obtain following expression for average current in spin channel $\sigma$
(for brevity of notation we suppress the superscript $\hat{\bm x}$ written in Eq.~(\ref{spcurr1}),
\begin{equation}
\langle I_{\alpha\hat{{\bm u}}}^{\sigma} \rangle =\frac{g}{h}\int_{0}^{\infty}{dE} \sum_{\beta}f_{\beta}(E)\left[N_{\alpha}^{\sigma}\delta_{\alpha\beta} - \sum_{\sigma^{\prime}mn}S^{\dagger\,\sigma^{\prime}\sigma}_{\beta m;\alpha n}
S^{\sigma\sigma^{\prime}}_{\alpha n;
\beta m}\right]
\label{spcurr2}
\end{equation}
Where $f_{\beta}={1/{exp[(E-\mu_{\beta})/kT]+1}}$ is Fermi distribution function with chemical potential $\mu_{\beta}$ and the pre-factor $g$ equals $\sigma\hbar/2$.
The summation over $\sigma^{\prime}$ in Eq.~(\ref{spcurr2}) can take on values $\pm\sigma$ corresponding to two spin projections along local spin quantization axis.
The second term of Eq.(\ref{spcurr2}) can be written explicitly in terms of spin resolved reflection and transmission probabilities as, 
\begin{eqnarray} 
\sum_{\beta\sigma^{\prime};mn}S^{\dagger \sigma^{\prime}\sigma}_{\beta m;\alpha n}
S^{\sigma\sigma^{\prime}}_{\alpha n;\beta m}&=&
\sum_{\sigma^{\prime};mn}S^{\dagger\,\sigma^{\prime}\sigma}_{\alpha m;\alpha n}
S^{\sigma\sigma^{\prime}}_{\alpha n;\alpha m} \nonumber \\ &+&\displaystyle\sum_{\beta\neq\alpha\sigma^{\prime};mn}S^{\dagger\,\sigma^{\prime}\sigma}_{\beta m;\alpha n}
S^{\sigma\sigma^{\prime}}_{\alpha n;\beta m} \nonumber \\
&\equiv& \displaystyle\sum_{\sigma^{\prime}}R_{\alpha\alpha}^{\sigma\sigma^{\prime}}+\displaystyle\sum_{\beta\neq\alpha\,\sigma^{\prime}}T_{\alpha\beta}^{\sigma\sigma^{\prime}}
\label{reftrans}
\end{eqnarray}
Where $R_{\alpha\alpha}^{\sigma\sigma^{\prime}}$ and $T_{\alpha\beta}^{\sigma\sigma^{\prime}}$ are spin resolved reflection and transmission probability in the same probe and between different probes respectively. In Eq.\ref{reftrans} on right hand side
spin resolved reflection and transmission
probabilities are summed over all possible input modes for a fixed output spin mode
$\sigma$ in lead $\alpha$. Because partial scattering matrix in spin subspace is not unitary due to non conservation of spin hence this summation need not to be equal to number of spin $\sigma$ channels in lead $\alpha$, {\it i.e.} $N_{\alpha}^{\sigma}$, rather it can have any value lying between zero and $N_{\alpha}^{\sigma}$. To
determine $N_{\alpha}^{\sigma}$ in terms of spin resolved reflection and transmission probabilities, consider a situation where current is injected from reservoir only in spin channels $\sigma$ in lead $\alpha$. In this case charge conservation requires that
this current should leave the spin channel $\sigma$ through all other possible channels in
the same lead as well in differing leads, which implies,
\begin{equation}
N_{\alpha}^{\sigma}=\sum_{\sigma^{\prime}}R_{\alpha\alpha}^{\sigma^{\prime}\sigma}+\sum_{\beta\neq\alpha\sigma^{\prime}}T_{\beta\alpha}^{\sigma^{\prime}\sigma}.
\label{chargecons}
\end{equation}
As we can see that Eq.(\ref{chargecons}) differs from Eq.(\ref{reftrans}) in a subtle way and are not equal because in general spin resolved transmission or reflection probabilities can not be related among themselves by interchanging spin indices, i.e., $T_{\alpha\beta}^{\sigma^{\prime}\sigma} \neq T_{\beta\alpha}^{\sigma\sigma^{\prime}}$ and
$R_{\alpha\alpha}^{\sigma -\sigma} \neq R_{\alpha\alpha}^{-\sigma \sigma}$ ( we will discuss
constraints due to time reversal symmetry below).
If we demand that sum in Eq.(\ref{reftrans}) also equals to $N_{\alpha}^{\sigma}$ then it would
imply spin conservation which is incorrect in presence of spin flip scattering or broken SU(2) symmetry. The inadvertent use of this charge conservation sum rule for spin degrees of freedom in Ref. \cite{schied,kiselev,jiang} has led to incorrect  spin current equation.
Though the partial scattering matrix in spin subspace is not unitary,however, the full scattering matrix is unitary,i.e., $SS^{\dagger}=S^{\dagger}S=I$, therefore, 
if we sum over $\sigma$ also in Eq.(\ref{reftrans}) or Eq.(\ref{chargecons})
then it should give total number of channels in leads $\alpha$, {\it i.e.} $N=N_{\alpha}^{\sigma}+N_{\alpha}^{-\sigma}$, and as a result we get the
following sum rule for total transmission probability, 
\begin{equation}
\mathcal{T_{\beta\alpha}}=\sum_{\sigma^{\prime}\sigma}T_{\beta\alpha}^{\sigma^{\prime}\sigma} =
\sum_{\sigma\sigma{\prime}}T_{\alpha\beta}^{\sigma\sigma^{\prime}}=\mathcal{T_{\alpha\beta}}
\label{sumrule}
\end{equation}
where $\mathcal{T_{\alpha\beta}}$ is total tranmission probability.

The net spin current flowing in lead $\alpha$ is defined as $I_{{\hat{\bm u}}\alpha}^S$=$ \langle I_{{\hat{\bm u}}\alpha}^{\sigma}\rangle +\langle I_{{\hat{\bm u}}\alpha}^{-\sigma}\rangle$ while the net charge current flowing is given by sum of absolute values, i.e., $I_{{\hat{\bm u}}\alpha}^q$=$ \mid \langle I_{{\hat{\bm u}}\alpha}^{\sigma}\rangle \mid+\mid\langle I_{{\hat{\bm u}}\alpha}^{-\sigma}\rangle\mid$ with pre-factor $g$ replace by the electronic charge $e$ in Eq.~(\ref{spcurr2}). Using Eqs.(
\ref{reftrans}),(\ref{chargecons}) in Eq.(\ref{spcurr2}) we obtain net spin and charge current as, 
\begin{eqnarray}
I_{\alpha\hat{\bm u}}^{s}=(\frac{\hbar }{2 h})\int_{0}^{\infty}{dE}\, \displaystyle
2 \,f_{\alpha}(E)(R_{\alpha\alpha}^{-\sigma\sigma}- R_{\alpha\alpha}^{\sigma -\sigma})+ \nonumber \\
\sum_{\beta\neq\alpha\sigma^{\prime}}
\left[ f_{\alpha}(E)(T_{\beta\alpha}^{\sigma^{\prime}\sigma} - T_{\beta\alpha}^{\sigma^{\prime}-\sigma}) %\right. \nonumber \\
%\left. 
-f_{\beta}(E)
(T_{\alpha\beta}^{\sigma\sigma^{\prime}} - T_{\alpha\beta}^{-\sigma\sigma^{\prime}}) \right]
\label{spchf}
\end{eqnarray}
\begin{eqnarray}
I_{\alpha\hat{\bm u}}^{q}=(\frac{e }{ h})\int_{0}^{\infty}{dE}\, \displaystyle
\sum_{\beta\neq\alpha\sigma^{\prime}}
\left[ f_{\alpha}(E)-f_{\beta}(E)\right]\mathcal{T_{\alpha\beta}}
\label{chcurr}
\end{eqnarray}
Equation (\ref{spchf}) is the central result of this work.{\it We stress that Eqs.~(\ref{spchf}) and (\ref{chcurr}) are valid under most general conditions as we have not made any assumptions about symmetries of the scattering region}.
It is instructive to note that in general $T_{\alpha\beta}^{\sigma^{\prime}\sigma}\neq T_{\alpha\beta}^{\sigma^{\prime}-\sigma}$ and
$R_{\alpha\alpha}^{\sigma -\sigma} \neq R_{\alpha\alpha}^{-\sigma \sigma} $ therefore, spin current equation can not be simplified further and written in terms of difference of Fermi function multiplied by transmission or reflection probabilities as is the case for charge current in Eq.~(\ref{chcurr}) which is standard Landauer-B\"uttiker result\cite{buttiker46}.
Hence the spin current given by Eq.~(\ref{spchf}) will be nonzero even when all the leads are at equilibrium, i.e., $f_{\alpha}(E,\mu_\alpha)=f(E,\mu),\forall \, \alpha$, where $\mu$ is equilibrium chemical potential. For sake of completeness we mention that equilibrium charge current vanishes as is evident from Eq.~(\ref{chcurr}). {\it The preceding discussion 
implies that
linear response for spin currents is not applicable in an electrical circuit where
external perturbation is applied voltages which is conjugate to charge currents and not to the
spin currents as discussed in introduction.}
Therefore,the most widely used equation for spin current,see Ref.\cite{schied,land-but} obtained by a generalization of charge current Eq.~(\ref{chcurr})
has to regarded as incorrect.
In view of this the theoretical study of spin dependent phenomena in mesoscopic systems needs
to be re-investigated.

We can gain further insight into spin current by considering non-equilibrium situation such that the chemical potential at the different leads differ only by a small amount so that we can expand the 
Fermi distribution function around equilibrium chemical potential $\mu$ as, $f_{\beta}(E,\mu_{\beta})=f(E,\mu)+(-df/dE)(\mu_{\beta}-\mu)$. In this case we can immediately notice from Eq.~(\ref{spchf}) that total spin current in non-equilibrium situation will have equilibrium as well non equilibrium parts of spin current. For ESC
the full Fermi sea of occupied levels will contribute. Therefore even in non-equilibrium situation
ESC cannot be neglected.
%For sake of completeness we recall that interlayer exchange coupling in magnetic multilayer systems arises due to ESC\cite{LevyFert}

{\it Equilibrium spin currents in time reversal symmetric two terminal system:}
In time reversal symmetric systems spin resolved transmission and reflection probabilities 
in Eq.~(\ref{spchf}) obey following relations {\it i.e.}, $R_{\alpha\alpha}^{\sigma\sigma^{\prime}}=R_{\alpha\alpha}^{-\sigma^{\prime}-\sigma}$ and 
$T_{\alpha\beta}^{\sigma\sigma^{\prime}}=T_{\beta\alpha}^{-\sigma^{\prime}-\sigma}$ \cite{pareek}.
In this case the spin currents Eq.~(\ref{spchf}) further simplifies to (here we denote left and right terminals by $L$ and $R$ respectively),
\begin{eqnarray}
I^{s,eq}_{L,\hat{\bm{u}}}=(\frac{\hbar }{2 h})\int_{0}^{\infty}{dE}\,2\,f(E,\mu)\left[ (R_{LL}^{-\sigma\sigma}-R_{LL}^{\sigma-\sigma})\right. \nonumber \\ \left.
+(T_{RL}^{\sigma-\sigma} - T_{RL}^{-\sigma\sigma})+(T_{RL}^{-\sigma-\sigma} - T_{RL}^{\sigma\sigma})\right],
\label{eqsptimerev}
\end{eqnarray}
above equation gives spin current in Left terminal. Spin current in right terminal are obtained from the same equation by interchanging $L \leftrightarrow R $. On right hand side in Eq.~(\ref{eqsptimerev}), $\sigma$ and $-\sigma$ refers to up and down spin states along
$\hat{\bm u}$.
From the above equation and previous discussion it is evident that
even in time reversal symmetric two terminal systems ESC are non zero.
Incase SU(2) symmetry in spin space is preserved, the spin resolved transmission and reflection probabilities obey a further rotational symmetry in spin space, i.e, $T_{\alpha\beta}^{\sigma\sigma^{\prime}}=T_{\alpha\beta}^{-\sigma-\sigma^{\prime}}$ ,
$R_{\alpha\alpha}^{\sigma\sigma^{\prime}}=R_{\alpha\alpha}^{-\sigma-\sigma^{\prime}}$ and spin flip components are zero, which implies that spin currents are identically zero for all terminals as is evident from Eq.~(\ref{eqsptimerev}).
This conclusion remains valid even for systems without time reversal symmetry as can be seen easily from Eq.~(\ref{spchf}).

The expression in Eq.~(\ref{eqsptimerev}) can be cast in a more useful form as (the details will be provided in Ref.\cite{paree_n}),
\begin{eqnarray}
I^{s,eq}_{\alpha {\hat {\bm u}}}=\frac{1}{2\pi}\int
Tr_{\sigma}\left[\{\Gamma_{\alpha}G^{r}\Gamma_{\beta}G^{a}+\right . \nonumber \\ \left.
(\Gamma_{\alpha}g^{r}\Gamma_{\alpha}g^{a})\}
(\bm{\sigma}\cdot{\bm u})\right]f(E,\mu)dE d{\bm k_{\parallel}}.
\label{traceformula}
\end{eqnarray}
In the above equation all symbols represents $2\times2$ matrices in spin space( in $\bm{\sigma}\cdot{\bm u}$ basis) and trace is taken over spin space. Where $\Gamma_{\alpha,\beta}$ represents broadening matrices due to contacts, $G^{r(a)}$ are retarded and advanced Green function and $g^{r,a}$ is
a off diagonal matrix in spin space defined as $g^{r,a}=[\{0,G^{r,a}_{\sigma-\sigma}\},\{G^{r,a}_{-\sigma\sigma},0\}]$. First term in Eq.~(\ref{traceformula}) corresponds to spin resolved transmission while the second and third term give spin resolved reflection probabilities as required by Eq.~(\ref{eqsptimerev}).
Notice that the above formula can not be simply written in terms of transmission matrix and it is reminiscent of the charge current formula for interacting system derived in Ref.\cite{meir}.
In our case this happens for spin current  because in presence of SO interaction spin can not be described as a non-interacting object.

{\it Equilibrium spin currents in two terminal Rashba system:}
We now apply Eq.(\ref{traceformula}) to study ESC(zero temperature) in a
finite size Rashba sample of length $L$, contacted by two ideal and identical unpolarized leads.
The Hamiltonian for two-dimensional electron system with Rashba SO interaction and short-range
spin independent disorder is $H=\hbar^{2}k^{2}/(2m^{*}) {\bm I}+\lambda_{so}(\sigma_{x}k_{y}-\sigma_{y}k_{x})+U(x,y)$, where $\lambda_{so}$ is Rashba SO coupling strength, $U(x,y)$ is the random disorder potential and $I$ is $2\times2$ identity matrix.
Neglecting weak localization effects, the disorder averaged retarded Green function including the effect of
leads is given by\cite{skvortsov},
\begin{equation}
G^{r,a}(E,\bm{k})=\frac{E-\frac{\hbar^2k^2}{2m}+i\eta(\bm k)+\lambda_{so}(\sigma_{x}k_{y}-\sigma_{y}k_{x})}{[(E-\frac{\hbar^{2}k^{2}}{2m^{*}}+i\eta(\bm k))^{2}-(\lambda_{so}k)^{2}]}
\end{equation}
with $\eta(\bm k)=(2\gamma(\bm k) +\frac{\hbar}{\tau(\bm k)})$ where $\tau(\bm k)$ is momentum relaxation time due to elastic scattering caused by impurities and
$\gamma(\bm k)$ broadening due to leads. The contact broadening matrices in Eq.~(\ref{traceformula}) are diagonal in spin space and defined as $ \Gamma_{1,2}=[\{\gamma(\bm k),0\},\{0,\gamma(\bm k)\}]$. Physically significance of $\gamma(k)$ is that it represents $\hbar/2$ times the rate at which an electron placed in a momentum state $k$ will escape into left lead or right lead, hence as a first approximation we can write, $\gamma(k)=\frac{\hbar v_{x}(\bm k)}{L}\equiv \frac{\hbar^2k\cos(\phi}{mL}$, where $\phi$ is angle with respect to $\bm x$ axis. The impurity scattering time can be approximated as $\frac{1}{\tau(k)}\approx \frac{\hbar k}{l_{el}}$, where $l_{el}$ is elastic mean free path.
With these inputs we can integrate Eq.~(\ref{traceformula}) over transverse
momentum (multichannel case) and energy to obtain an analytical expression for equilibrium spin current. We find that ESC with spin parallel(antiparallel) to the ${\hat{\bm z}}$ or ${\hat{\bm z}}$ axis and flowing to the {\it x} direction vanishes in both leads($I^{\hat{\bm x} s}_{\hat{\bm z}}\equiv \sigma_{z}v_{x}$=0, $I^{\hat{\bm x} s}_{\hat{\bm x}}\equiv\sigma_{x}v_{x}$=0). The ESC with spin parallel(antiparallel) to the $\hat{\bm y}$ axis are nonzero and given by,
\begin{eqnarray}
I^{-\hat{\bm x},s}_{L,+\hat{\bm y}}=I^{+\hat{\bm x},s}_{R,-\hat{\bm y}}\approxeq {\frac{m^{*}\lambda_{so}E_{F}L}{32\pi \hbar^{2}}}(2+(\frac{L}{l_{el}})^{2}).
\label{eqrashba}
\end{eqnarray}
In left lead ESC is polarized along $+\hat{\bm y}$ direction and flows outwards from sample to lead,i.e.,along $-\hat{\bm x}$ direction, while in the right lead ESC is polarized along $-\hat{\bm y}$ and flows outwards 
from sample to lead,i.e.,along $+\hat{\bm x}$ direction.
Physically this implies that 
spin angular momentum is generated in sample with SO coupling which then flows outwards in the
regions where SO coupling is zero, i.e., the left and right leads. {\it This implies a spin rectification effect which can only occur if the transport is non linear and we see that this consistent with nonlinear nature of spin currents as remarked earlier.}
It is important to note that
due to ESC there is no net magnetization in the total system (sample+leads) which is consistent with the Kramer's degeneracy.

We can gain a deeper understanding of the above expression if we
analyze the systems using additional symmetries. The disorder
averaging establishes reflection symmetry with respect to
{\it x(y)} axis and the system has a symmetry related with the operator $\sigma_{y}R_{y}$. ($\sigma_{x}R_{x}$). As a result the total symmetry operator(time reversal+reflection) for the system is
$\mathcal{U_{TR}}=I_{t}\sigma_{x}R_{x}\sigma_{y}R_{y}=I_{t}(i\sigma_{z})R_{x}R_{y}$, where $I_{t}$ is time reversal operator. Under this symmetry operation the disorder averaged system is invariant
and the spin current operators $\sigma_{x}v_{x}$, $\sigma_{y}v_{x}$ are even while $\sigma_{z}v_{x}$ is odd. Therefore the spin current along $\hat{\bm z}$ direction vanishes
while in-plane spin current can be non zero. However as we have seen above that only the
$\bm y$ component of spin current survives after integrating over momentum.
Fig.(1) illustrate the conservation of $\sigma_{y}v_{x}$ under these symmetry operation.
The dependence of ESC on $\lambda_{so}$ can also be inferred from symmetry consideration.
As we can check easily that the Rashba SO interaction changes sign under reflection along $\hat{\bm z}$ axis($\lambda_{so}(\sigma_{x}k_{y}-\sigma_{y}k_{x}) \mapsto -\lambda_{so}(\sigma_{x}k_{y}-\sigma_{y}k_{x})$). Physically it corresponds to reversing the
asymmetry of confining potential along $\hat{\bm z}$ axis. Therefore the spin currents (equilibrium as well non-equilibrium) can only depend on the odd powers of spin orbit coupling constants $\lambda_{so}$. According to Eq.~(\ref{eqrashba}) ESC are proportional to
$\lambda_{so}$ which is consistent with these
symmetry consideration and ESC vanishes
if $\lambda_{so}$ zero as expected on physical grounds. 
It is worth noting that even the local ESC in
macroscopic Rashba medium, discussed in Ref.\cite{rashba} are proportional to $\lambda_{so}^{3}$ and as we have seen this is a consequence of symmetry consideration.
Moreover ESC are proportional to length of SO region because SO region acts as source of 
these currents. Note that the right hand-side of Eq.~(\ref{eqrashba}) has dimension of angular momentum per unit time signifying that these currents are truly transport current.

\begin{figure}
\includegraphics[width=\linewidth,height=1.8in,angle=0,keepaspectratio]{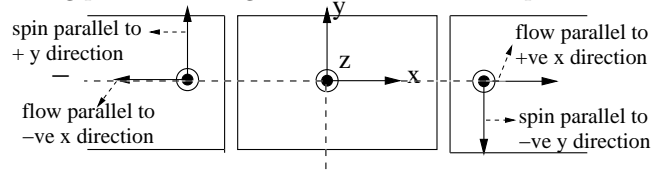}
\caption{Fig 1: The figure illustrate the conservation of spin current $\sigma_{y}v_{x}$
if th system is rotated by $\pi$ along $\hat{\bm z}$ axis, which represents reflection along and y axis respectively. Under this transformation configuration in left and right goes over into each other hence the spin current remains invariant.}
\label{Fig. 1}
\end{figure}
To conclude, we have derived spin current formula for multiterminal spin transport
for system with broken SU(2) symmetry in spin space. We have demonstrated that spin currents are
fundamentally different from the charge currents and the ESC are generically non zero. In view of this the spin transport phenomena in mesoscopic system 
needs a fresh look. Further it will also be interesting and desirable to study different
magnetoresistance phenomena from the perspective of spin currents.

I acknowledge helpful discussion with M. B\"uttiker and A. M. Jayannavar.

\end{document}